\documentclass[%
 reprint,
 amsmath,amssymb,
 aps,
]{revtex4-2}

\usepackage[utf8]{inputenc}
\usepackage{braket}

\usepackage{graphicx}
\usepackage{dcolumn}
\usepackage{bm}
\usepackage{hyperref}
\usepackage{balance}

\usepackage{mathrsfs}

\begin{document}

\title{Quantum Fourier Addition, Simplified to Toffoli Addition}

\author{Alexandru Paler}
\affiliation{
    Aalto University, Espoo 02150, Finland
}
\affiliation{
    University of Texas at Dallas, Richardson, TX 75080, USA
}

\email{alexandrupaler@gmail.com}

\begin{abstract}
Quantum addition circuits are considered being of two types: 1) Toffolli-adder circuits which use only classical reversible gates (CNOT and Toffoli), and 2) QFT-adder circuits based on the quantum Fourier transformation. We present the first systematic translation of the QFT-addition circuit into a Toffoli-based adder. This result shows that QFT-addition has fundamentally the same fault-tolerance cost (e.g. T-count) as the most cost-efficient Toffoli-adder: instead of using approximate decompositions of the gates from the QFT circuit, it is more efficient to merge gates. In order to achieve this, we formulated novel circuit identities for multi-controlled gates and apply the identities algorithmically. The employed techniques can be used to automate quantum circuit optimisation heuristics.
\end{abstract}

\maketitle

\section{Introduction}

Addition circuits are one of the workhorses of quantum circuit design. The literature on arithmetic circuits includes two types of adders \cite{rines2018high, ruiz2017quantum}: QFT-adders and Toffoli-based adders. Intuitively, it is the same addition algorithm (one in a  Boolean space, the other in a phase space), but there is a gap in the interpretation of one addition circuit in terms of the other. For an excellent introduction to the basics of quantum arithmetic circuits we refer the reader to \cite{rines2018high}. The challenge we are solving herein is to show the systematic translation between the two types of addition circuits.

Recent applications of QFT-adders are quantum walks \cite{walks} for NISQ computers. Nevertheless, while QFT-adders are considered by some to be NISQ-compatible due to their depth and gate counts, there are works on Toffoli-based arithmetic for NISQ \cite{thapliyal2021quantum}. Recently, approximate addition has been proposed in \cite{approxadder}, which can be seen as a mix between QFT and Clifford+T due to the gate set those circuits are using.

\subsection{Motivation}

Most of the metrics that influence the fault-tolerant implementation of quantum computations are related to circuit depth, number of wires, T-count. The cost of quantum addition has been investigated from different perspectives, but for practical applications it has almost always boiled down those metrics. Fault-tolerant circuits, even if initially expressed using Toffoli gates, have to be compiled to Clifford+T gates because, practical codes, such as the surface code, can only protect this gate set.

Yoder~\cite{yoder2017universal}, for example, presented fault-tolerant Toffoli-like gates using Bacon-Shor codes, but such constructions are not compatible with the surface code. For this reason, resource estimations of practical circuits just replaced the QFT-addition with the more cost effective reversible one (e.g. \cite{haner2017factoring}).

Our contribution is to show that for arithmetic circuits, the QFT can be avoided by systematically applying circuit identities from \cite{barenco1995elementary}. We effectively translate a QFT-adder into a Toffoli-based adder. Independent of this work, efforts \cite{kuijpers2019graphical} to connect the ZX-calculus (used for Clifford+T circuits) to the ZH-calculus (used for Toffoli + H circuits) have highlighted that the connection is mediated by the QFT. The authors of \cite{kuijpers2019graphical} analysed the cost of addition without looking at explicit addition circuits, but used the QFT to derive generalised Toffoli gates. From this perspective, our result is a special case of the result from \cite{kuijpers2019graphical} (it goes back from a single QFT to multiple Toffoli gates), as well the first explicit translation between QFT- and Toffoli-adders.

\subsection{Background}

\begin{figure}
    \centering
    \includegraphics[width=\columnwidth]{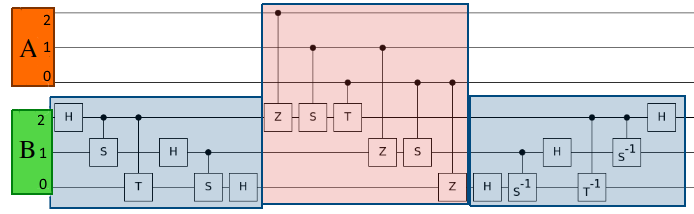}
    \caption{A 3 qubit QFT-adder for adding the numbers A and B. The three qubit register A is orange, and the three-qubit register B is green. The blue regions are the QFTs (direct $QFT$ and inverse $QFT^\dagger$). The red region includes the controlled rotations.}
    \label{fig:init}
\end{figure}

For the purpose of this work, we assume that in a QFT-adder (e.g. Fig.~\ref{fig:init}), the numbers $A$ and $B$ are encoded in two multi-qubit registers: the addition is implemented by applying the QFT on register $B$, then systematically rotating by well specified angles the relative phase of the individual qubits from the register of $B$, and finally undoing the QFT on register $B$. The rotations are controlled by the $A$ register.

Logical circuits for reversible addition \footnote{in the sense of Boolean reversible logic} can be expressed exclusively with three-qubit Toffoli and two-qubit CNOT gates. This fact has the advantage that reversible adder constructions can be inspired by classical computer architecture, and some examples are the ripple carry \cite{cuccaro2004new} and the carry save adders \cite{gossett1998quantum}. Once addition circuits are formulated with reversible gates, there are many exact or empirical methods to optimise the reversible circuits with respect to a given cost function.

One of the disadvantages of the fault-tolerant compilation of QFT-adders to Clifford+T is that the controlled rotations in the QFT-adders are approximated by CNOTs and sequences of single qubit gates using algorithms such as \cite{gridsynth}, which incur a non-negligible T-count and T-depth overhead to the resulting circuit. On the other hand, the fault-tolerant versions of Toffoli-adders (such as the ripple carry adder also-known-as the Cuccaro adder \cite{cuccaro2004new}) have a T-count linear in the number of Toffoli gates whose number is linear in the number of the adder wires.

\section{Results}

\begin{figure}
    \centering
    \includegraphics[width=0.7\columnwidth]{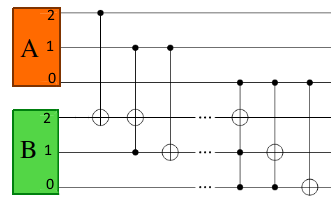}
    \caption{A QFT-adder expressed using multi-qubit Toffoli gates. Similarly to the initial circuit, the number of controlled operations by the qubits from A is maintained. For example, the third qubit from A was controlling three rotations in the QFT case, and now it is part of three Toffoli gates.}
    \label{fig:final}
\end{figure}

Very often it is assumed that a QFT is not related to Toffoli-based circuits. In this work, we translate a quantum Fourier adder circuit into a Toffoli based adder. The construction is iterative, and by induction can be applied to adders of arbitrary width. In Section~\ref{sec:meth} we detail the steps necessary to obtain a reversible adder (Fig.~\ref{fig:final}) from an initial QFT-adder (Fig.~\ref{fig:init}). To simplify terminology, we consider all n-qubit-controlled applications of the $X$ ($Z$) gate as a generalised ($n+1$)-qubit Toffoli (CCZ) gate. The CNOT (CZ) is a two-qubit Toffoli (CCZ) gate.

Every Toffoli-based adder has a regular structure. This holds also for the adder we derive herein: it performs bitwise addition starting from the highest bits (top wires in each register), after having considered the potential carry bits from all the lower sums. For example, the left-most gate controlled by $A_2$ in Fig.~\ref{fig:final} is computing the XOR-sum $A_2 \oplus B_2$, without accounting for carry bits. Afterwards the next two gates (a Toffoli and a CNOT) are being controlled by $A_1$: the first gate updates the value of $B_2$ to account for the carry arriving from $A_1 \oplus B_1$, and the second gate is effectively computing the XOR-sum $A_1 \oplus B_1$. This gate pattern is continued for the entire addition circuit.

Reversible addition circuits have a Toffoli gate count which is linear in the number of addition bits. Thus, it is advantageous to compile Toffoli-based adders and then to decompose each generalised Toffoli gate into sequences of three-qubit Toffoli gates~\cite{barenco1995elementary}. The T-count of the decomposed Toffoli-based addition circuits will be linear in the number of qubits of the adder, whereas when the rotations are approximated the T-count is a function of approximation and scales logarithmically with the inverse of the approximation error \cite{gridsynth}. For small angles in medium sized QFT-addition circuits (e.g. 128 qubits) the T-count can easily reach millions.

Systematically transforming a QFT-adder into a reversible one uses a sequence of steps which are repeated until the resulting circuit consists entirely of Toffoli gates. The result of our procedure is a circuit that has a regular gate structure (Fig.~\ref{fig:final}). A second systematic application of gate rewrite rules, based on Toffoli gate commutations, could be applied to obtain the ripple carry adder from \cite{cuccaro2004new}. Our reversible adder and the ripple carry adder share a common property: they require no ancillae.

\section{Methods}
\label{sec:meth}

In this section we will: a) explain the systematic transformation steps (insert QFT and its inverse, and merge/fuse rotation gates) from a QFT- to a Toffoli-based adder, b) present an example of how rotation gates are merged, and c) list the transformation algorithm. A step-by-step transformation of a three qubit QFT-adder is illustrated in the Appendix.

\begin{figure}[h!]
    \centering
    \includegraphics[width=0.85\columnwidth]{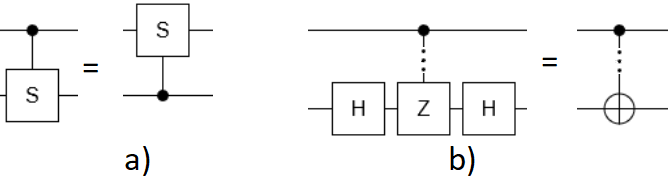}
    \caption{a) The direction of controlled rotations around the Z axis can be changed. In this example, the controlled-S gate. b) Hadamard gates are merged into n-qubit Toffoli gates (in this example CNOT).}
    \label{fig:cz}
\end{figure}

The first circuit identity we use (Fig.~\ref{fig:cz}a) states that the direction of controlled rotation gates around the Z axis can be flipped. We will also need to remove Hadamard gates (Fig.~\ref{fig:cz}b) from the circuits by merging these with the target of multi-qubit CCZ gates to obtain Toffoli gates.

\subsection{Inserting QFT and its inverse}
\label{sec:qft}

\begin{figure}[!h]
    \centering
    \includegraphics[width=\columnwidth]{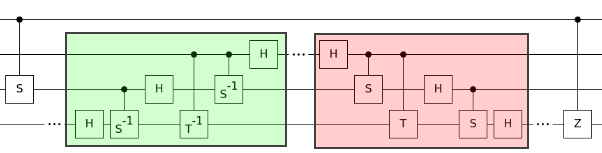}
    \caption{Inserting a QFT (green, left) and its inverse (red, right) between two controlled rotations. This leaves the computation unchanged because the QFTs cancel. The controlled rotation gates will be used to form larger angle rotations until multi-qubit CCZ gates are obtained. The newly inserted Hadamards will be merged together with existing Hadamard gates (not illustrated) and multi-qubit CCZ gates to form multi-qubit Toffoli gates.}
    \label{fig:insqft}
\end{figure}

The method starts by inserting a reverse ($QFT^\dagger$) and a direct QFT (Fig.~\ref{fig:insqft}). The QFTs are inserted before the CZ controlled by the lowest bit from the $A$ register ($A_0$). The Hadamard gates from the two QFTs transform the CZ into a CX (like in Fig.~\ref{fig:cz}). The next step is to group rotations of the same angle around the CNOT such that the identity from Fig.~\ref{fig:sqrt} is applied. Systematic gate cancellations and application of the same identities, as explained in the following section, return the reversible adder from Fig.~\ref{fig:final}.

\subsection{Merging rotation gates}

\begin{figure}[h!]
    \centering
    \includegraphics[width=0.7\columnwidth]{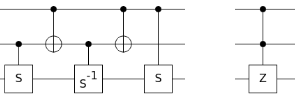}
    \caption{Implementing a CZ with controlled-S gates. The circuit identity holds in general for $V^2=U$. In this example $V=S$ and $U=Z$.}
    \label{fig:sqrt}
\end{figure}

The QFT and the QFT-adder include small angle controlled rotations which we merge (fuse) into rotations of larger angles. We employ the reverse of a circuit identity, in the following called SQRT, which is the decomposition of arbitrary unitaries into their square roots. For the of example of $S^2=Z$, where $S$ is the square root of the $Z$ gate, the controlled application of $Z$ can be expressed using the circuit from Fig.~\ref{fig:sqrt}\cite{barenco1995elementary}.

\begin{figure}[h!]
    \centering
    \includegraphics[width=0.7\columnwidth]{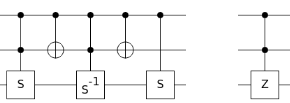}
    \caption{Equivalent circuit for a double controlled unitary $Z$: adding second controls to the first two controlled-S gates does not change the computation.}
    \label{fig:sqrt2}
\end{figure}

\begin{figure}[h!]
    \centering
    \includegraphics[width=\columnwidth]{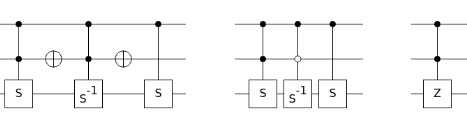}
    \caption{Removing the controls from the CNOTs does not change the computation. Two double controlled-$S$ gates and single controlled-$S$ gate to implement a double controlled $Z$.}
    \label{fig:sqrt3}
\end{figure}

Novel equivalent formulations of SQRT are presented in Fig.~\ref{fig:sqrt2} and \ref{fig:sqrt3}. The correctness of the circuit identities can be shown by computing the exponents of the unitaries that are being applied in a controlled manner (an approach similar to the one from Section 7 in \cite{barenco1995elementary}).

It is possible to apply SQRT to multiple-controlled operations, as shown in the following. The generalisation is enabled by how the CNOTs are applied in Fig.~\ref{fig:sqrt}. The first step is to double the control wires of the controlled-S gate (Fig.~\ref{fig:sqrt2}).

The previous circuit transformation allows us now to transform the CNOTs into simple XORs (Fig.~\ref{fig:sqrt3} for the example of $V=S$ and $U=Z$).

\begin{figure}[h!]
    \centering
    \includegraphics[width=0.7\columnwidth]{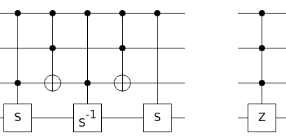}
    \caption{Multi-qubit controlled gate, where $U=Z$ using $V=S$.}
    \label{fig:sqrt4}
\end{figure}

Finally, it is possible using Fig.~\ref{fig:sqrt} and \ref{fig:sqrt2}, to generalise the SQRT rule to multi-qubit controlled operations (Fig.~\ref{fig:sqrt4}).

\subsection{Example of merging rotations}
\begin{figure}[h!]
    \centering
    \includegraphics[width=0.7\columnwidth]{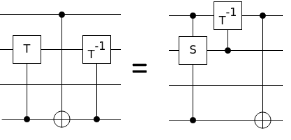}
    \caption{Example of using the SQRT identity as a step during the translation algorithm. The left hand side is a sub-circuit appearing in the QFT-adder. The right hand side is the result of applying the SQRT rule.}
    \label{fig:example00}
\end{figure}

The example from Fig.~\ref{fig:example00} will arise in a three-qubit QFT-adder, after inserting a QFT and its inverse, and merging two Hadamard gates with a CZ to form a CNOT.

\begin{figure}[h!]
    \centering
    \includegraphics[width=0.6\columnwidth]{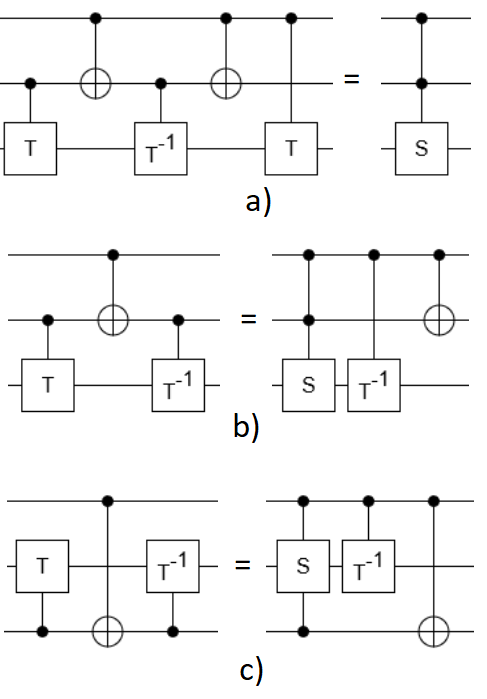}
    \caption{Example illustrating the correctness of the transformation from Fig.~\ref{fig:example00}. a) The SQRT rule for T and S gates; b) a CNOT and controlled-T are commuted on the side of the double-controlled-S gate; c) the second and third wire are swapped.}
    \label{fig:example01}
\end{figure}

In the following we show step-by-step how to derive the result from Fig.~\ref{fig:example00}. We start from the circuit from Fig.~\ref{fig:sqrt2} where we replace S with T and Z with S (Fig.~\ref{fig:example01}a). Afterwards, we move the CNOT and controlled-T gate on the side of the double-controlled-S gate (Fig.~\ref{fig:example01}b). Finally, we permute the wires and, up to the application of the identity from Fig.~\ref{fig:cz}a, we obtain a circuit equivalent to the right hand side of Fig.~\ref{fig:example00}.

\subsection{Algorithm}
\label{sec:alg}
The idea of the QFT-adder simplification is to use the SQRT identity to combine pairs of $m$-qubit controlled rotations around the Z axis with an angle of $\frac{\pi}{2^p}$ into single $(m+1)$-qubit controlled rotation with an angle of $\frac{\pi}{2^{p-1}}$. For the circuit from Fig.~\ref{fig:init} the controlled-T uses the maximum value $p=3$.

The first part of the translation algorithm is to insert a pair of QFTs (Section~\ref{sec:qft}). The second part of the algorithm has complexity $\mathcal{O}(n^4)$, where $n$ is the number of qubits (the QFT has $\mathcal{O}(n^2)$ gates, and the algorithm is repeated for each gate to find a matching pair to apply the SQRT rule). The second part of the algorithm is repeating the following steps until the circuit has only Toffoli gates:
\begin{enumerate}
    \item If possible, eliminate pairs of Hadamard gates surrounding the control of a multi-qubit CCZ and transform into a multi-qubit Toffoli gate (Fig.~\ref{fig:cz}b);
    \item Commute pairs of opposite angle rotations (e.g. T and T$^\dagger$) next to the new multi-qubit Toffoli;
    \item Flip the direction of the rotations (Fig.~\ref{fig:cz}a);
    \item Apply SQRT to the pair of rotations and the CNOT. Obtain a new CNOT and controlled rotation (Fig.~\ref{fig:example00});
    \item Find the left-most controlled rotation that cancels the newly introduced rotation.
\end{enumerate}

\section{Discussion and Conclusion}

We showed how to transform a QFT-adder into a Toffoli-adder, by effectively eliminating the QFT from the addition circuit. Instead of decomposing the controlled rotations from the QFT-adder, the rotations are merged into ones with larger angles until multi-qubit Toffoli gates are obtained. 

Eliminating the QFT from computations (e.g. \cite{aaronson2020quantum}) has been recently of practical interest, because, although the circuits have shallow depth in the presence of gate parallelism, the small angles make it difficult for NISQ devices. Small angle rotations are a significant issue also for the resource efficiency of error-corrected quantum circuits, such that approximate QFT circuits have been proposed and recently optimized \cite{nam2020approximate}.

Moreover, it seems that, in some situations, the QFT is not a significant component of quantum circuits (e.g. \cite{aharonov2006quantum, van2013efficient}). For example, the Toffoli+H gate set is universal \cite{shi2003both}, and the ZX and ZH calculus have equivalent computational power \cite{kuijpers2019graphical}. Additionally, eliminating the QFT from circuits might have implications also on quantum circuit simulation methods.

QFT-adders are not fundamentally different from the reversible adders. This opens the possibility of optimising QFT-based circuits systematically, and also to compile QFT-Toffoli hybrid circuits.

\section*{Acknowledgements}

We thank Daniel Herr for correcting a first version of the circuit transformations. A.P. was supported by a Google Faculty Research Award and a Fulbright Senior Researcher Fellowship.

\bibliography{__main}

\section*{Appendix}

\begin{figure}[h!]
    \centering
    \includegraphics[width=\columnwidth]{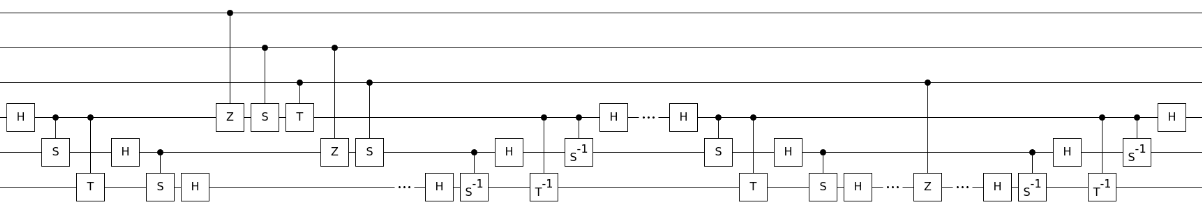}
    \caption{An inverse and a direct QFT are inserted before the right-most CZ gate.}
    \label{fig:op02}
\end{figure}

\begin{figure}[h!]
    \centering
    \includegraphics[width=\columnwidth]{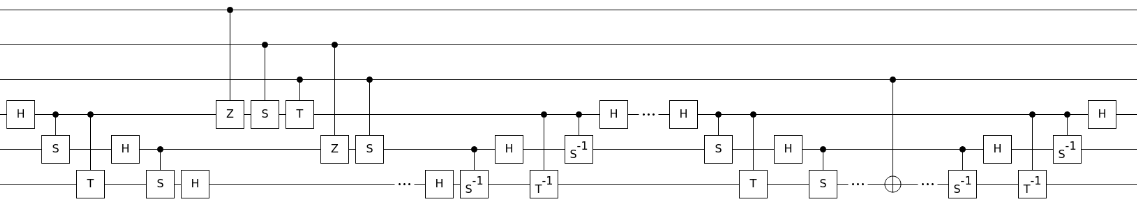}
    \caption{The two Hadamards surrounding the CZ gate transform it into a CNOT.}
    \label{fig:op03}
\end{figure}

\begin{figure}[h!]
    \centering
    \includegraphics[width=\columnwidth]{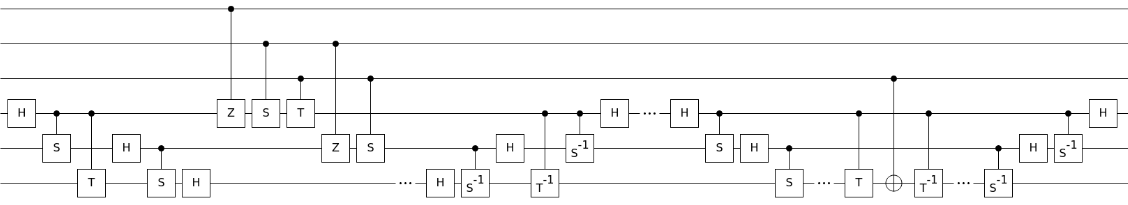}
    \caption{The controlled-T and controlled-T$^\dagger$ are commuted right next to the CNOT.}
    \label{fig:op04}
\end{figure}

\begin{figure}[h!]
    \centering
    \includegraphics[width=\columnwidth]{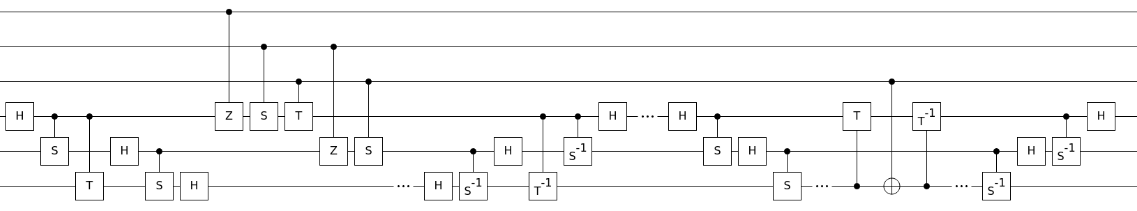}
    \caption{The direction of the controlled-T gates is changed.}
    \label{fig:op05}
\end{figure}

\begin{figure}[h!]
    \centering
    \includegraphics[width=\columnwidth]{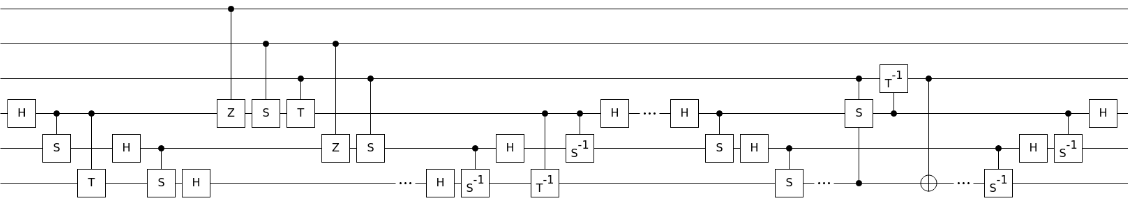}
    \caption{The SQRT rule is applied. The result are double-controlled-S gate, a CNOT and a controlled-T$^\dagger$.}
    \label{fig:op06}
\end{figure}

\begin{figure}[h!]
    \centering
    \includegraphics[width=\columnwidth]{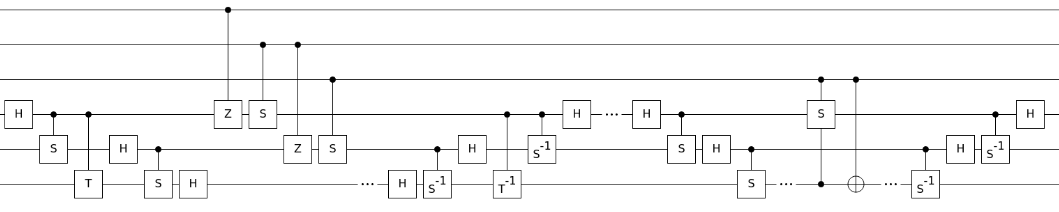}
    \caption{The freshly introduced controlled-T$^\dagger$ cancels with the left-most controlled-T.}
    \label{fig:op07}
\end{figure}

\begin{figure}[h!]
    \centering
    \includegraphics[width=\columnwidth]{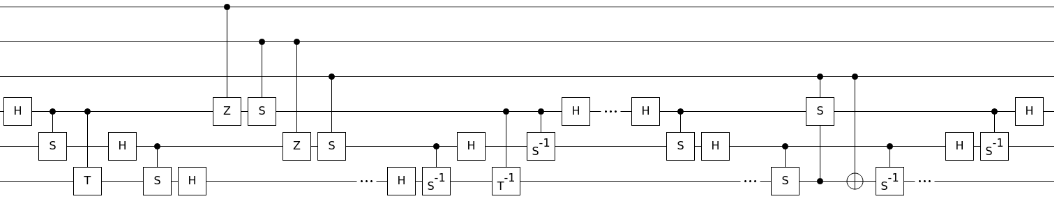}
    \caption{The controlled-S and controlled-S$^\dagger$ are commuted right next to the CNOT.}
    \label{fig:op08}
\end{figure}

\begin{figure}[h!]
    \centering
    \includegraphics[width=\columnwidth]{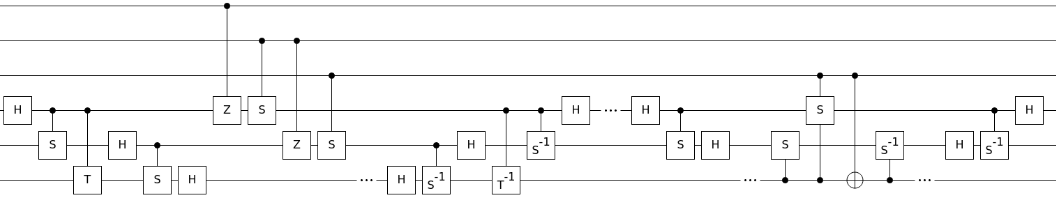}
    \caption{The direction of the controlled-S gates is changed.}
    \label{fig:op09}
\end{figure}

\begin{figure}[h!]
    \centering
    \includegraphics[width=\columnwidth]{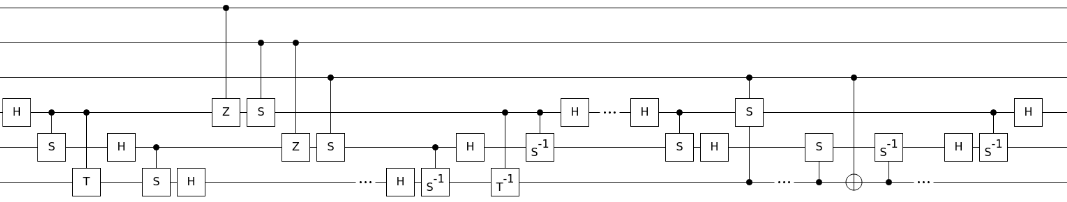}
    \caption{An equivalent circuit to Fig.~\ref{fig:op09}}
    \label{fig:op10}
\end{figure}

\begin{figure}[h!]
    \centering
    \includegraphics[width=\columnwidth]{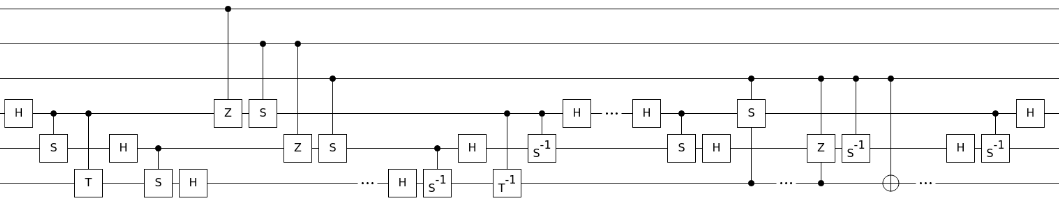}
    \caption{The SQRT rule is applied. The result are a CCZ gate, a CNOT and a controlled-S$^\dagger$.}
    \label{fig:op11}
\end{figure}

\begin{figure}[h!]
    \centering
    \includegraphics[width=\columnwidth]{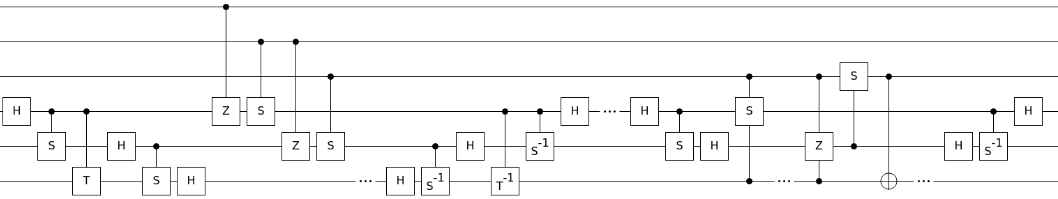}
    \caption{The direction of the freshly introduced controlled-S$^\dagger$ is changed.}
    \label{fig:op12}
\end{figure}

\begin{figure}[h!]
    \centering
    \includegraphics[width=\columnwidth]{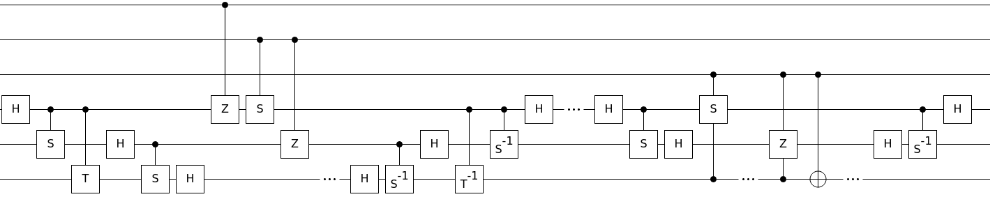}
    \caption{The freshly introduced controlled-S$^\dagger$ cancels with the left-most controlled-S.}
    \label{fig:op13}
\end{figure}

\begin{figure}[h!]
    \centering
    \includegraphics[width=\columnwidth]{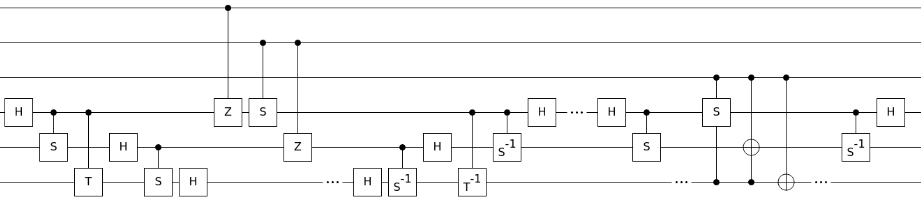}
    \caption{The Hadamards surrounding the CCZ transform the gate into a Toffoli.}
    \label{fig:op14}
\end{figure}

\begin{figure}[h!]
    \centering
    \includegraphics[width=\columnwidth]{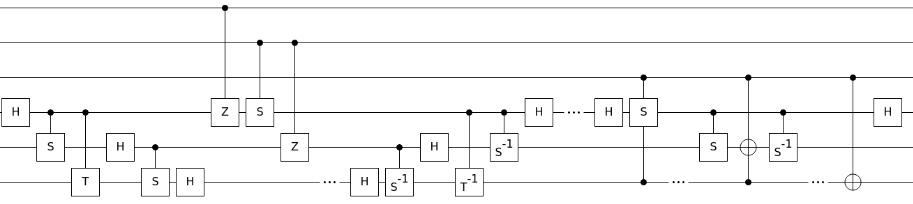}
    \caption{The controlled-S and controlled-S$^\dagger$ are commuted right next to the CNOT.}
    \label{fig:op15}
\end{figure}

\begin{figure}[h!]
    \centering
    \includegraphics[width=\columnwidth]{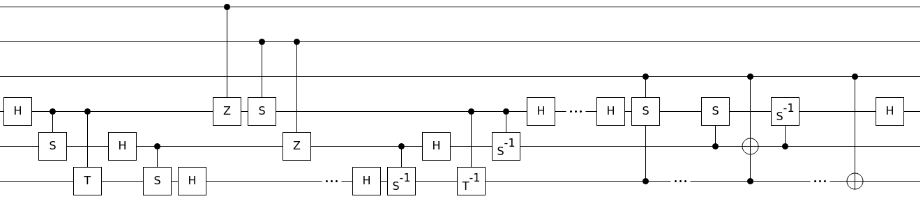}
    \caption{The direction of the controlled-S gates is changed.}
    \label{fig:op16}
\end{figure}

\begin{figure}[h!]
    \centering
    \includegraphics[width=\columnwidth]{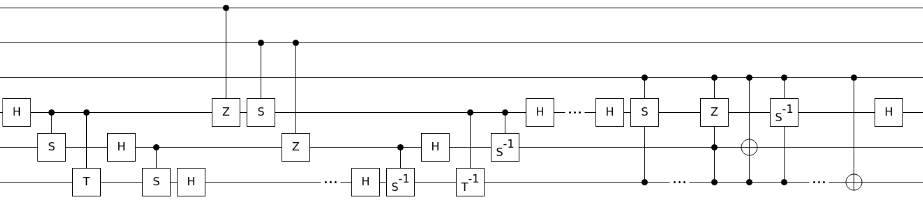}
    \caption{The SQRT rule is applied. The result are a four-qubit-controlled Z gate, a Toffoli and a double-controlled-S$^\dagger$.}
    \label{fig:op17}
\end{figure}

\begin{figure}[h!]
    \centering
    \includegraphics[width=\columnwidth]{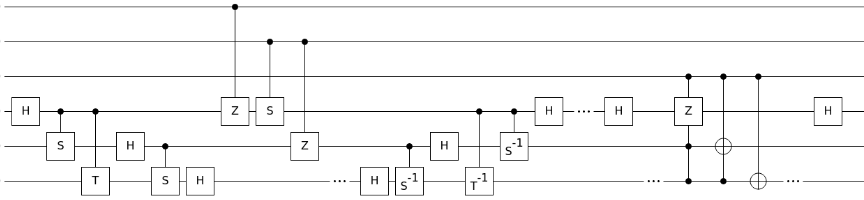}
    \caption{The double-controlled-S$^\dagger$ cancels with the left-most double-controlled-S.}
    \label{fig:op18}
\end{figure}

\begin{figure}[h!]
    \centering
    \includegraphics[width=\columnwidth]{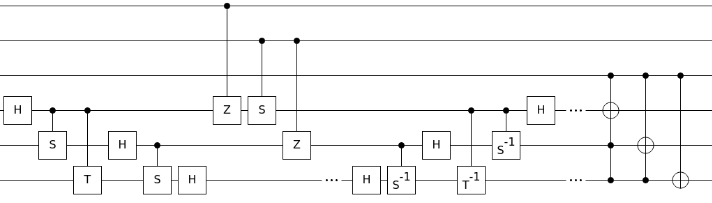}
    \caption{The Hadamards surrounding the CCCZ transform the gate into a four-qubit Toffoli.}
    \label{fig:op19}
\end{figure}

\begin{figure}[h!]
    \centering
    \includegraphics[width=\columnwidth]{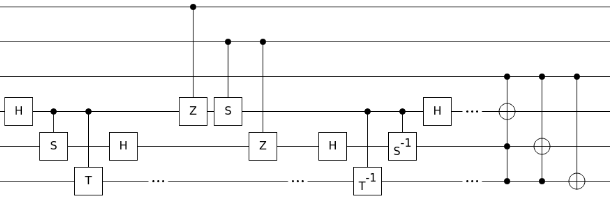}
    \caption{The Hadamards and the controlled-S gates on the bottom wire cancel.}
    \label{fig:op21}
\end{figure}

\begin{figure}[h!]
    \centering
    \includegraphics[width=0.75\columnwidth]{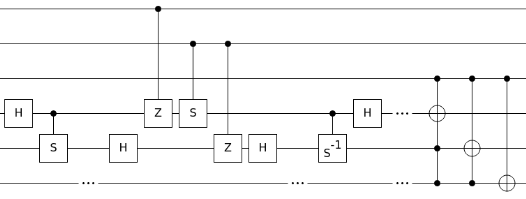}
    \caption{The controlled-T gates on the bottom wire cancel.}
    \label{fig:op23}
\end{figure}

\begin{figure}[h!]
    \centering
    \includegraphics[width=0.75\columnwidth]{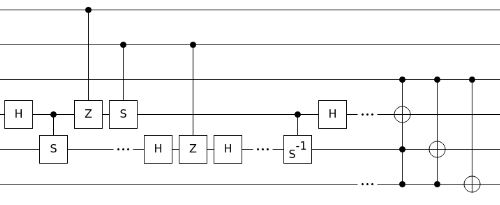}
    \caption{The two Hadamards are moved next to the CZ.}
    \label{fig:op24}
\end{figure}

\begin{figure}[h!]
    \centering
    \includegraphics[width=0.75\columnwidth]{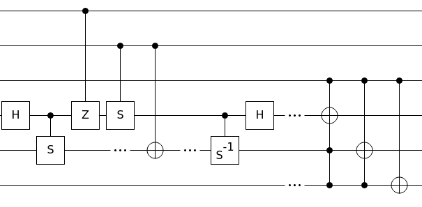}
    \caption{The two Hadamards surrounding the CZ gate transform it into a CNOT.}
    \label{fig:op25}
\end{figure}

\begin{figure}[h!]
    \centering
    \includegraphics[width=0.75\columnwidth]{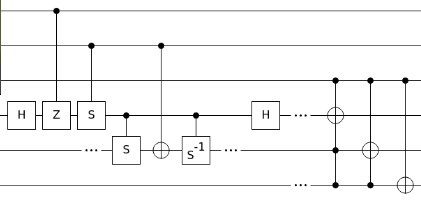}
    \caption{The controlled-S and controlled-S$^\dagger$ are commuted right next to the CNOT.}
    \label{fig:op26}
\end{figure}

\begin{figure}[h!]
    \centering
    \includegraphics[width=0.75\columnwidth]{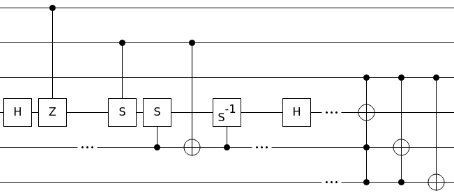}
    \caption{The direction of the controlled-S gates is changed.}
    \label{fig:op27}
\end{figure}

\begin{figure}[h!]
    \centering
    \includegraphics[width=0.75\columnwidth]{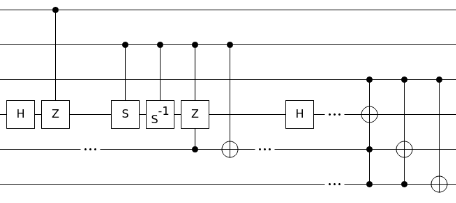}
    \caption{The SQRT rule is applied. The result are a CCZ gate, a CNOT and a controlled-S$^\dagger$.}
    \label{fig:op28}
\end{figure}

\begin{figure}[h!]
    \centering
    \includegraphics[width=0.75\columnwidth]{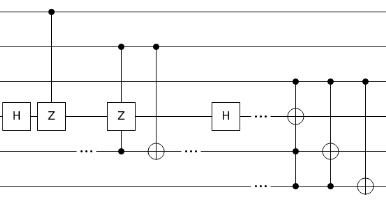}
    \caption{The freshly introduced controlled-S$^\dagger$ cancels with the left-most controlled-S.}
    \label{fig:op29}
\end{figure}

\begin{figure}[h!]
    \centering
    \includegraphics[width=0.75\columnwidth]{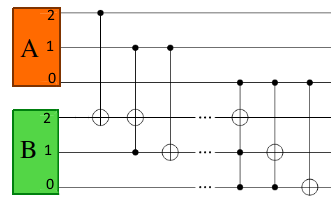}
    \caption{The Hadamards surrounding the targets of the multiple-controlled Z gates change the gates into Toffolis.}
    \label{fig:op30}
\end{figure}

Figs.~\ref{fig:op02}--\ref{fig:op30} illustrate the step-by-step transformation of a three-qubit QFT-adder into a Toffoli-adder. We include these for completeness. Wider addition circuits can be transformed similarly by applying the algorithm sketched in Section~\ref{sec:alg}.

\clearpage

\end{document}